\documentclass[]{aa}

\usepackage{graphicx}
\usepackage{txfonts}
\usepackage{natbib}
\bibpunct{(}{)}{;}{a}{}{,}

\def\pcite#1{[ref]}

\begin{document}
   \title{Total mass distributions of Sersic galaxies from photometry $\&$ central velocity dispersion}
   \subtitle{}
   \titlerunning{Mass Distributions Only from Photometry $\&$ $\sigma_0$}
   \author{Dalia Chakrabarty \inst{1}
           \and Brendan Jackson  \inst{2}
          }
   \offprints{Dalia Chakrabarty}
   \institute{
              School of Physics $\&$ Astronomy,
              University of Nottingham, 
              Nottingham NG7 2RD, U.K.
              \email{dalia.chakrabarty$@$nottingham.ac.uk}
              \and 
              Institute for Astronomy, 
              University of Edinburgh, 
              Blackford Hill, 
              Edinburgh EH9 3HJ, UK
              \email{bmj@roe.ac.uk}
             }

   \date{\today}

\abstract {} {We develop a novel way of finding total mass
density profiles in Sersic ellipticals, to about 3 times the major axis
effective radius, using no other information other than what is
typically available for distant galaxies, namely the observed surface
brightness distribution and the central velocity dispersion
$\sigma_0$.} {The luminosity density profile of the observed
galaxy is extracted by deprojecting the measured brightness
distribution and scaling it by a fiduciary, step-function shaped, $raw$
mass-to-light ratio profile ($M/L$). The resulting raw, discontinuous, total,
3-D mass density profile is then smoothed according to a proposed smoothing
prescription. The parameters of this raw $M/L$ are characterised by
implementing the observables in a model-based study.} {The
complete characterisation of the formalism is provided as a function
of the measurements of the brightness distribution and $\sigma_0$. The
formalism, thus specified, is demonstrated to yield the mass density
profiles of a suite of test galaxies and is successfully applied to
extract the gravitational mass distribution in NGC~3379 and NGC~4499, 
out to about 3 effective radii.}{}


\maketitle

\section{Introduction}
\noindent
Any evaluation of the total mass in distant galaxies is a struggle
against the paucity of available observational evidence. Photometry is
hardly enough to indicate the content of both the luminous as well as the
dark matter, unless the functional dependence between luminosity
content and dark mass is accessible. This is of course not the case;
the existence of such a relation is itself uncertain,
especially in early type galaxies. While kinematic information of
tracers has often been advanced as indicators of mass distributions in
galaxies, the implementation of such information is tricky, primarily
because of the mass-anisotropy degeneracy. Thus, one often resorts to
cleverly designed observational techniques and/or algorithms and
formalisms which thrive even in light of the limited measurements.
Examples of these include the Planetary Nebula Spectrograph
\citep{roma_science}, the NMAGIC code \citep{nmagic_07} and CHASSIS
\citep{saha_dal}. 

The X-ray emission from X-ray active systems can be
analysed to offer insight into the gravitational mass distribution in
the galaxy, under the assumption of hydrostatic equilibrium \citep[to
cite some recent work]{humphrey_08, lemze_08, mahdavi_08, zhang_07,
fukazawa_06, ewan_04}. However, what makes this method potentially
unreliable is the lack of information about the distribution of the
fraction of hot gas that is in hydrostatic equilibrium
\citep{churazov_08, diehl_07}.

Comparatively, a more stable route to mass distribution determination
is via lensing measurements. However, the biggest shortcoming of mass
determination from lensing measurements alone is the unavailability
of the full three-dimensional mass distribution. To improve upon this,
lensing data is often supplemented by dynamically obtained mass
estimates \citep{czoske_08, bolton_08, gavazzi_07, kt_03}.

However, there are questionable implementational problems involved in
(parametric) dynamical mass determination, the chief of which
are typically the mass-anisotropy degeneracy, binning-triggered
instability of scant velocity dispersion data, reliance on the
modelling of the stellar mass density and an even more fundamental
worry caused by the assumption of one smooth parametric representation
of the phase-space distribution function of the used tracer and by its
relation to the phase-space density of the whole galaxy. This is of
course in addition to the uncertainties in the mass distribution
resulting from substituting the real geometry of the system by
sphericity, as is typically done with all mass determination
procedures. Above all, using tracer kinematics for mass
determination \citep[etc.]{nmagic_08, douglas_07} is limited in
applicability given the reliance on the size of the tracer kinematic
data! Large data sets are of course hard to attain in systems that are
not close by. Moreover, this method is unsuitable for fast evaluation of
the mass distribution of individual galaxies that are members of a
large sample, as for example, a galaxy obseved in a large survey.

On the contrary, it would be highly beneficial to design a method that
is comparatively less data-intensive in that it demands only what is
easily available from observations. We advance a methodology that
provides total gravitational mass density distributions to about 3
effective radii, as compared to only 1 effective radius
\citep{sauron_06}, in a fast and easy-to-implement fashion.

This advanced formalism is inspired by a trick that was 
reported in \cite{chakrabarty07} (hereafter, Paper~I). This trick
involves the exploitation of only photometry and the central velocity
dispersion measure ($\sigma_0$) in a galaxy, in order to generate the
total local mass-to-light ratio ($M/L$) profile to a distance that is
about {\it thrice} the semi-major axis effective radius. This cutoff
distance is described in details below. The prescription for 
constructing this profile was provided in Paper~I, though only for
a certain class of power-law galaxies. However, the exact nature of
this prescription is very much a function of the photometric class
that the galaxy belongs to. Thus, the formula reported in Paper~I
cannot be invoked to shed light on the mass distribution in galaxies
that betray a different (and more ubiquitous) photometric class,
eg. ellipticals, the surface brightness of which can be fit by a
Sersic profile \citep{sersic_68}. This is precisely what is reported
in this paper.

This paper is arranged as follows: the basic framework of the
suggested formalism is discussed in Section~2, followed by a note on
the models that we use. The method used to obtain the sought
functional forms is briefly mentioned in Section~4. Results obtained
from our work are subsequently discussed in Section~5. Tests of the
method are described in Section~6 while Section~7 deals with
applications to real galaxies NGC~3379 and NGC~4494. The paper is
rounded off with a section devoted to discussions of relevant points.

\section{Formalism}
\begin{figure}
\includegraphics[width=8cm]{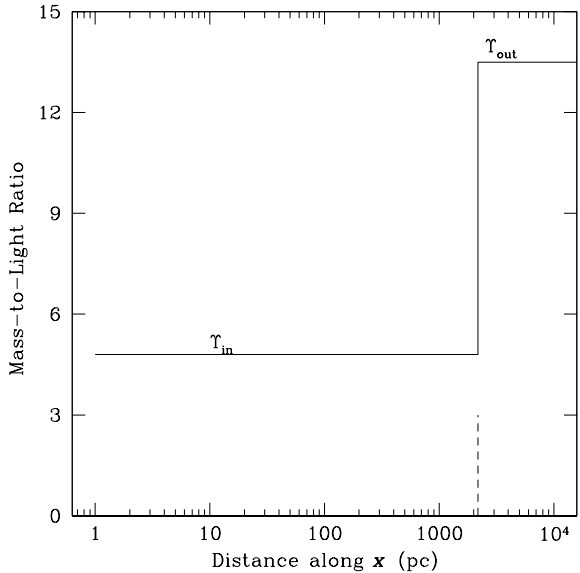}
\caption{\label{fig:scheme} Schematic view of a raw $M/L$ distribution
along the major axis coordinate $x$, for a model galaxy. The dashed
line marks the position of the jump radius $x_{in}$ which as defined
in Equation~\ref{eqn:xe}, is 3 times the major axis effective radius of the
model galaxy, if the galaxy is described by a Sersic index of 4;
otherwise, $x_{in}$ is an approximation for the major-axis effective
radius. The amplitudes inside and outside $X_{in}$ have been marked as
$\Upsilon_{in}$ and $\Upsilon_{out}$. While the measured central
velocity dispersion is used to estimate the range of $\Upsilon_{in}$
values allowed, $\Upsilon_{out}$ is constrained via the sought
analytical relation that connects $\Upsilon_{out}$ to
$\Upsilon_{in}$. The point of this paper is to seek such relations, in
order to fully characterise the raw $M/L$ profile; once the raw $M/L$
distribution is known, it is smoothed to obtain the final, total $M/L$
ratio profile of the galaxy, which is significantly different from
this unprocessed $M/L$ shown above. }
\end{figure}

\noindent
The only Sersic model that was considered in Paper~I did actually
indicate that the mass estimation trick suggested for the power-law
systems might be possible for Sersic galaxies too. Following this
lead, as in Paper~I, we first invoke a raw two-stepped $M/L$ profile
of the Sersic galaxy at hand, where it is the distribution of $M/L$
with the major axis coordinate $x$, that is relevant. This raw $M/L$
distribution is subsequently smoothed, (according to the smoothing
prescription provided in Paper~I and discussed below) to provide the
real $M/L$ distribution of the system, to a distance that is by
definition, 3 times the major-axis effective radii for a Sersic
galaxy with sersic index $n$=4 but is an approximation for the
major-axis effective radius for all other values of $n$. The formal
definition of this distance is given in Equation~\ref{eqn:xe} while
the justification for our choice of this length scale is delineated in
Section~\ref{sec:X_e}. Figure~\ref{fig:scheme} represents a schematic
diagram of this raw $M/L$ profile against $x$.

As can be appreciated from Figure~\ref{fig:scheme}, the unprocessed
$M/L$ profile is a two-stepped function that is characterised by three
free parameters, including the position of the step, or the jump
radius $x_{in}$ and the $M/L$ amplitudes inside and outside $x_{in}$.
The object of the current exercise is to fully characterise this
unprocessed $M/L$ profile, so that upon smoothing, the final, total,
local $M/L$ distribution over $x$ is retrieved. This final form of the
$M/L$ distribution is significantly different from the discontinuous,
2-stepped, raw $M/L$ that we initially choose to work with. In fact,
the final form of the profile is smooth and the $M/L$ values vary a
lot with $x$, sometimes abruptly, depending on the details of the
model galaxy. Figure~4 in Paper~I represents a comparison of one such
raw $M/L$ profile and the final smoothed $M/L$ distribution that is
advanced as the representative $M/L$ profile for the test system at
hand. In fact, later in Figure~\ref{fig:compare_ml}, a raw $M/L$
(exemplified in Figure~\ref{fig:scheme}) is compared to the true $M/L$
distribution of the test galaxy under consideration - the difference
between the raw and final forms of the $M/L$ profiles is clear in that
figure.

As in Paper~I, $x_{in}$ is set equal to 3$X_e$, where
\begin{equation}
X_e=\displaystyle{\left(\frac{-3.33}{m}\right)^4}
\label{eqn:xe}
\end{equation}
and $m$ is the slope of the straight line that is fit to the plot of
core-removed log$_{10}$(I) against $x^{1/4}$ (as in Paper~I),
i.e. $X_e$ is the major axis equivalent of the effective radius, if
the Sersic index is 4. For all other values of the Sersic index, $X_e$
is at most an approximation to the major axis effective radius. The
advantage of using $X_e$ over the exact definition of the major-axis
effective radius is described below in Section~\ref{sec:X_e}.

The smoothing prescription used here is the same as in Paper~I - we
smooth the raw total mass density profile by two successive
applications of a box filter of size corresponding to $X_e$.

The amplitude of $M/L(x)$ for $x \leq x_{in}$ is referred to as
$\Upsilon_{in}$ and that for $x > x_{in}$ it is $\Upsilon_{out}$. Now
that $x_{in}$ has been pinned down by construction, we hope to get a
constraint on the available choices for $\Upsilon_{in}$ from the
central velocity dispersion ($\sigma_0$) and also hope to identify a
functional dependence of $\Upsilon_{out}$ on $\Upsilon_{in}$.  The
exact form of such a constraint or function is yet unknown but we
begin by expecting these to be defined in terms of the photometric
parameters that describe the surface brightness profile of the Sersic
galaxy under consideration, namely, the Sersic index $n$ and $X_e$
that we have defined above (Equation~\ref{eqn:xe}). The central
brightness is not a free parameter since we normalise all luminosity
density profiles to a central value of 1000
L$_{\odot}$pc$^{-3}$. Thus, we want to find the function $f(n, X_e,
\Upsilon_{in})$, where
\begin{equation}
\Upsilon_{out} = f(n, X_e, \Upsilon_{in})
\end{equation}

The inspiration for the hypothesis that $f$ depends on the photometric
parameters is discussed below.


Given that by construction, $\Upsilon_{in}$ is the uniform amplitude
of the raw $M/L$ profile for $x < x_{in}$, we expect it to be related
to the ``central'' mass-to-light ratio of the galaxy, where by
``central'' is implied the distance at which the measurement of the
central velocity dispersion ($\sigma_0$) is obtained (at $x=x_0$, with
$x_0$ typically less than $x_{in}$). However, in a real system, the
true central $M/L$ cannot be securely determined from the measurement
of $\sigma_0$ alone, owing to uncertainties about the validity of the
assumptions that are invoked, in order to translate knowledge of
$\sigma_0$ to that of mass enclosed within $x_0$ (using virial
theorem). Such uncertainties basically stem from the presence of
anisotropy in phase-space. In the mass modelling trick advanced in
Paper~I, room is allowed for the accommodation of such uncertainties,
as long as the deviations from the assumptions used in the virial
estimate of mass are not atypically more than what has been observed
with real ellipticals \citep{padmanabhan04}; this is discussed in the
following paragraph.

In Paper~I, the virial estimate of the central $M/L$, from $\sigma_0$
was parametrised by $\alpha$. It was found that for a measured
$\sigma_0$, as long as $\Upsilon_{in}$ lies within a range of values
(the details of this range correspond to the given $\alpha$),
compatibility between the predicted and known (model) mass density
distributions is ensured. In other words, for $\alpha$ calculated from
a given $\sigma_0$, $\Upsilon_{in}$ can be safely chosen to belong to
a range of $M/L$ values: $\Upsilon_{in}^{min}$ to
$\Upsilon_{out}^{max}$. Such positioning of $\Upsilon_{in}$ can be
checked by comparing the mass distributions recovered with
$\Upsilon_{in}=\Upsilon_{in}^{min}$ and
$\Upsilon_{in}=\Upsilon_{in}^{max}$, for consistency. Here
$\Upsilon_{in}^{min}=\alpha$ while $\Upsilon_{in}^{max}$ is an unknown
function of the photometric parameters and $\alpha$ - say $g(n, X_e,
\alpha)$. 
\begin{equation}
\Upsilon_{in}^{max} = g(n, X_e, \alpha),
\end{equation}
where $g(n, X_e, \alpha)$ is unknown and the choice of its dependence
on the photometric parameters is the following. In Paper~I we had
success upon choosing the functions $f$ and $g$ as dependent on the
photometric details of the system; this was the case for a suite of
model power-law galaxies.  Such ``success'' is qualified in terms of
the identification of the hypothesised dependence on the photometric
properties of the model galaxies.  Motivated by this, we endeavour to
find forms of $f$ and $g$.

Thus, there are two unknown functions that we wish to constrain: $f(n,
X_e, \Upsilon_{in})$ and $g(n, X_e, \alpha)$. These functions, when
known, will provide $\Upsilon_{out}$ from $\Upsilon_{in}$ which will
be known from $\alpha$, (i.e. $\sigma_0$) and the observed brightness
distribution of the galaxy. Once we know $\Upsilon_{out}$ from
$\Upsilon_{in}$, we would then have fully characterised the raw $M/L$
distribution over $x$. The smoothed out version of this raw $M/L$
profile will then be advanced as the true $M/L$ profile of the system
at hand. This in conjunction with the luminosity density distribution
will allow knowledge of the mass density distribution.

We hope to recover the analytical forms of these two unknown functions
for Sersic galaxies to 3$X_e$, through an analysis of a sample of
model Sersic galaxy surface brightness profiles. Analytical fits are
sought to the data that is collated from the suite of models that we
work with, in order to recognise patterns, if any, that may show up in
the relations between the various quantities, in particular,
$\Upsilon_{out}$-$\Upsilon_{in}$ and $\Upsilon_{in}-\alpha$. The
variation in these relations with changing models is then explored to
unravel the reliance of these relations on the photometric parameters.

\begin{figure*}
\includegraphics[width=16.5cm]{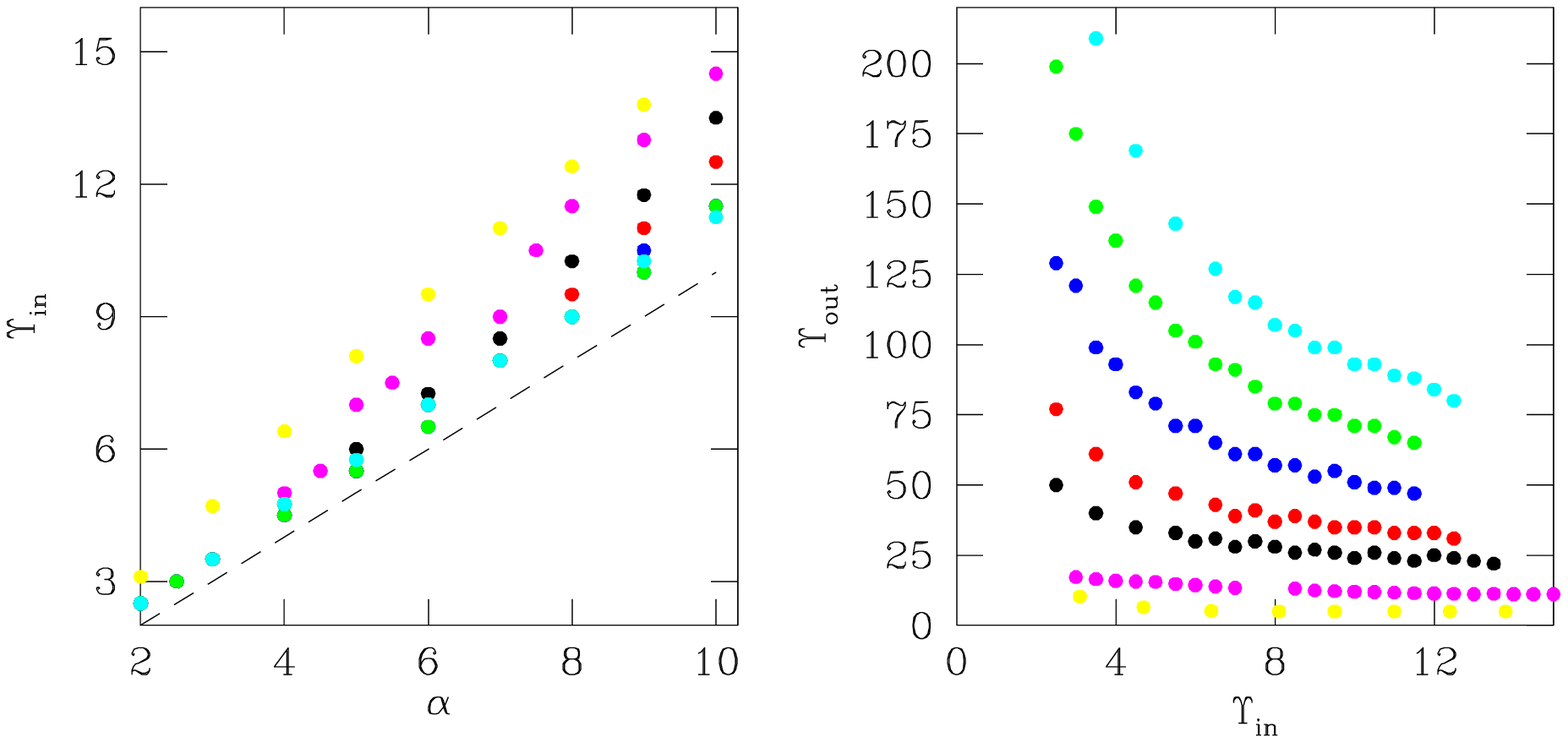}
\caption{\small{Right panel: trends in $\Upsilon_{out}$ with
$\Upsilon_{in}$, with changing Sersic index $n$, from model galaxies
with $X_e$=726 pc. The colour coding used for the different $n$ values
is: yellow for $n$=3, magenta for $n$=4, black for $n$=5.3, red for
$n$=6, blue for $n$=7, green for $n$=8 and cyan for $n$=9. In the left
panel, $\Upsilon_{in}^{min}$ is plotted against $\alpha$ in broken
lines; this plot is universally given by
$\Upsilon_{in}^{min}=\alpha$. $\Upsilon_{in}^{max}$ is plotted against
$\alpha$ for the different model galaxies used for the figure on the
right, with the same colour coding.  For a model at hand, at a given
$\alpha$, $\Upsilon_{in}$ can be chosen from the range defined by the
values of $\Upsilon_{in}^{min}$ and $\Upsilon_{in}^{max}$. }}
\end{figure*}

\section{Models}
\label{sec:models}
\noindent
We identify the relation between $\Upsilon_{in}$ and $\Upsilon_{out}$
in Sersic galaxies, by recognising (and then quantifying) the patterns
that emerge between these quantities, as diverse model galaxies are
scanned. To reminisce, Sersic galaxies are those, the surface
brightness profile $I(R)$ of which can be approximated as:
\begin{equation}
I(R) = I_0\exp\left[-b_n(R/R_e)^{1/n}\right],
\end{equation}
where $I_0$ is the central surface brightness, $R_e$ is the projected
effective half-light radius, $n$ is the Sersic index that determines
the curvature of the brightness profile and $b_n$ is a function of
$n$: for $n > 0.5, b_n \approx 2n - 1/3 + 0.009876/n$
\citep{prugniel97, lima_99}. Thus, the Sersic model represents a
3-parameter family. The corresponding deprojected spherical luminosity
density distribution is discussed in \cite{terzic05, mazure_02}. In
\cite{terzic07}, the spherical models of \cite{terzic05} were
generalised to include triaxiality. The luminosity density profile in
the triaxial case is approximately given as:
\begin{eqnarray}
\label{eqn:density}
\rho_L(s) &=& \rho_0\displaystyle{\left({\frac{R_e}{s}}\right)^p e^{-b_n(s/R_e)^{1/n}}} \quad \textrm{where} \\ 
p &=& \displaystyle{1.0 - \frac{0.6097}{n} + \frac{0.05563}{n^2}} \quad \textrm{for $n\in(0.6, 10)$}. \nonumber
\end{eqnarray}
Here, the equation for $p$ is due to \cite{lima_99} and the
ellipsoidal coordinate $s$ is defined as
\begin{equation}
s^2 \equiv \displaystyle{\frac{x^2}{a^2}+\frac{y^2}{b^2} + \frac{z^2}{c^2}}.
\end{equation}
for the axial ratios of a:b:c for the ellipsoidal system at hand. We
use $X_e$ in place of $R_e$ (see Section~7.5). It is to be noted that
Equation~\ref{eqn:density} is an approximation and other forms have
been used by \cite{nacho_02}.

As in Paper~I, in our models, we assume oblateness and an inclination
of 90$^{\circ}$, with a uniform projected axial ratio of 0.7; the work
of \cite{padilla_08} corroborates such considerations of geometry and
ellipticity. Also, the model systems are assumed to be viewed in the
ACS $z$-band, at a distance of 17 Mpc. Though we recover the sought
functions for these chosen configurations, generalisations to other
systems will be suggested in Section~\ref{sec:photo} and
Section~\ref{sec:D}.

Actually \cite{terzic07} gives the more general 5-parameter mass
density model which describes Sersic galaxies with a core inside a
given break radius. However, we work with the simpler case of Sersic
galaxies that can be qualified by a 3-parameter model. In fact, we
constrict this further, by normalising the central luminosity density
$\rho_0$ to 1000 $M_\odot pc^{-3}$. Also, we look for the luminosity
distribution along the major axis. Thus, our models are distinguished
only by the Sersic index $n$ and the $X_e$ defined in equation~\ref{eqn:xe}.



Models were formulated for Sersic indices $n$=3, 4, 5.3, 6, 7, 8 and 9
with $X_e$ of 726 pc, 1000 pc, 1400 pc, 2000 pc. While these models
were extensively explored to achieve the sought functions, we
investigated selected models with $n$ lying in this range and higher
$X_e$ values. In particular, we are interested in models with Sersic
index $\leq$4, with larger $X_e$. Our viable models include galaxies
with $n$=3, $X_e\leq$4 kpc and $n$=4, $X_e\leq$5 kpc. Such an extended
size range holds special interest for the greater fraction of observed
ellipticals that correspond to this low $n$. Additionally, models with
other $n$ were also found to work to $X_e$ = 10 kpc. This choice of
models is supported by the results that
\begin{itemize}
\item in the nearby Virgo cluster, the ACSVCS survey \citep{laura_acs}
reports that nearly 90$\%$ of the targetted early-type systems (Sersic
and cored-Sersic included) fall in the range of $R_{e}\lesssim$5 kpc
in the $z$-band (which is the waveband directly comparable to our
models). All the programme galaxies with $n\in$[3,4] were found to
have $R_{e}\leq$ 2.2 kpc while the single galaxy with $n\in$(4,5.3]
and $R_e>$5 kpc, is not a Sersic galaxy but is a cored Sersic
system. Here, we remind ourselves that the reported effective radius
$R_e$ is really the geometric mean of the extent along the photometric
semi-axes. Thus, the extent along the semi-major axis is greater (by
about a factor of about 1.2, for an axial ratio of 0.7) than the
reported $R_e$. In other words, the surrogate for $X_e$ is about 1.2
times the values of $R_e$ quoted earlier in this paragraph. Even when
this factor in taken into account, the range of our models covers the
programme galaxies of the ACSVCS.

\item measurement of effective radius is waveband dependent
\citep{temi_08, ko_05}, so that for $n$=3 and 4, our limits on the
model $X_e$ values in the $z$-band, are compatible with observations
\citep{laura_acs, nacho_01, barbera_05}.

\item at high redshifts, systems display evolution towards higher
compactness \citep{fernando_08}. Though this evolution is marked for
systems at $z>$1.7 \citep[sample of][]{fernando_08},
\cite{ferreras_09} suggest a milder size evolution for the most
massive galaxies for 0.4$\leq z \leq$1.2. Thus, in the higher redshift
range that we propose our scheme to be most useful in, if anything,
the size range of our models would be more suitable.

\end{itemize} 
This corroborates our choice of models, particularly in regard to the
values of $X_e$.

Sersic indices 2 or lower appeared not to be viable for the formalism
to function, implying that this formalism is suitable only for
elliptical systems (see Section~\ref{sec:models}). Sersic indices
greater than 9 were not examined since such systems are very rare.

These luminosity density models, described by
Equation~\ref{eqn:density} and the used values of $n$ and $X_e$, were
embedded in an NFW-type dark halo \citep{NFW} of mass density
$\rho_{dark}$, to give a total mass density of:
\begin{equation}
\rho_t(x) = \rho_{dark}(x) + \alpha\rho_L(x),
\end{equation}
where by choice,
\begin{equation}
\rho_{dark}(x) = \displaystyle{\frac{M_s}{4{\pi}x(x+r_s)^2}}.
\end{equation}
Here $M_s$ and $r_s$ are mass and length scales of the halo,
respectively. Our results are valid for halo parameters that
correspond to the points in the green quadrilateral in the $M_s-r_s$
space that is depicted in the right panel of Figure~\ref{fig:DM}.

We ascribe $\pm$1-$\sigma$ errors of about 10$\%$ to the luminosity
density distributions that we generate and search for compatibility
between the model mass density distribution and the predicted one,
within these error bars.

\section{Method}
\noindent
Combinations of the parameters $\alpha$, $\Upsilon_{in}$ and
$\Upsilon_{out}$ that showed compatability with the known (model) mass
density profiles, up to 3$X_e$, at the aforementioned four separate
$(M_s,r_s)$ coordinates, were searched for by our smoothing formalism
that is automated. 

We record the list of $\Upsilon_{in}$ values that imply compatibility
for a chosen $\alpha$ (chosen typically in the range of 2 to 10). The
starting value of $\Upsilon_{in}$ was typically about $\alpha$ while
the upper value was set as 3$\alpha$, which was always sufficient to
find the whole range of $\Upsilon_{in}$ values that correspond to
compatability. Similarly, we record the $\Upsilon_{out}$ value
corresponding to a given $\Upsilon_{in}$.

The plots of $\Upsilon_{in}^{min}$ and $\Upsilon_{in}^{max}$, as
functions of $\alpha$ are monitored, with the aim of recognising the
analytical form of these functional dependences. Similarly, the plot
of $\Upsilon_{out}$ as a function of $\Upsilon_{in}$ is analysed at
different $n$ and $X_e$, to identify $f(n, X_e, \Upsilon_{in})$.

\section{Results}
\noindent
With $X_e$ held at 726pc, as $n$ is varied, the plots of
$\Upsilon_{in}$ against $\alpha$ are shown in the left panel of
Figure~2 while the relation between $\Upsilon_{out}$ against
$\Upsilon_{in}$ is depicted for these models, in the right panel of
this figure.

\subsection{Relation between $\Upsilon_{in}$ $\&$ $\alpha$}
\label{sec:upin_upout}
\noindent
When the data collated from the different models is plotted it is
noted that while $\Upsilon_{in}^{min}=\alpha$, the
$\log(\Upsilon_{in}^{max})-\log(\alpha)$ relation is well fit by a
quadratic function of $\log(\alpha)$. This latter functional form has
only a weak bearing on the Sersic index $n$, though
$\Upsilon_{in}^{max}$ is found to increase slightly as $n$ is
increased from 5.3 to 9. Thus, by denying this increased amplitude of
$\log\Upsilon_{in}^{max}$ with higher values of $n$, we merely
constrict the range of values from which $\Upsilon_{in}$ can be
chosen. Thus, our attempt at simplification of the sought functional
form preempts a small reduction in the applicability of our formalism
to more anisotropic galaxies than what is effectively allowed.

\begin{figure}
\includegraphics[width=8cm]{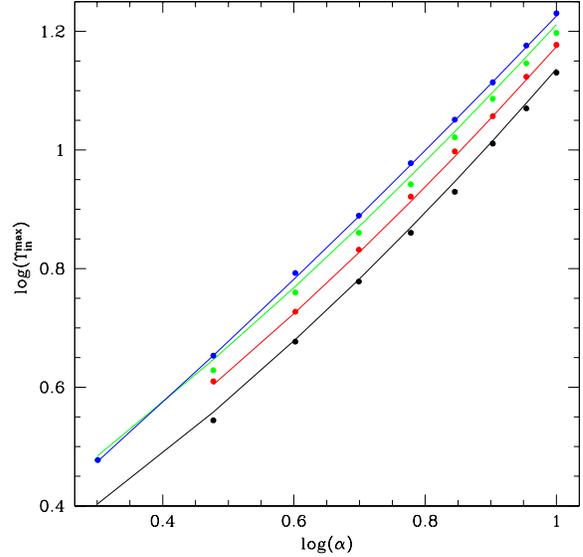}
\caption{\label{fig:alpha} The values of $\log\Upsilon_{in}^{max}$
obtained at distinct values of $\log\alpha$, from our models, for
$n$=5.3 and different $X_e$; in black filled circles for $X_e$=726 pc,
red for $X_e$=1000 pc, green for $X_e$=1400 pc and blue for $X_e$=2000
pc. The predicted dependence of $\log\Upsilon_{in}^{max}$ on
$\log\alpha$ (Eqn.~\ref{eqn:alpha_in}) is plotted in solid lines, for
the 4 different values of $X_e$ that we use, in corresponding colours.
This functional form holds approximately for other values of $n$ as
well ($n \geq$ 3); the amplitude of $\Upsilon_{in}^{max}$ only
increases slightly with $n$. }
\end{figure}

In Figure~\ref{fig:alpha}, we present the values of
$\log(\Upsilon_{in}^{max})$ at distinct values of $\log(\alpha)$, for
different $X_e$, and a single Sersic index of 5.3. The functional form
of the relation is recovered from our analysis, and this is
over-plotted on the data, in solid lines. This functional form is given
by:
\begin{eqnarray}
\label{eqn:alpha_in}
\Upsilon_{in}^{max} &=& A_0(X_e) + A_1(X_e)\log\alpha + A_2(X_e)(\log\alpha)^2\\ \nonumber
A_0(X_e) &=& -0.00747849 + 0.000360389X_e \\ \nonumber
& &-1.31416\times10^{-7}X_e^2 \\ \nonumber
A_1(X_e) &=& 0.597653 -2.61661\times10^{-5}X_e \\ \nonumber
& &+9.34449\times10^{-8}X_e^2 \\ \nonumber
A_2(X_e) &=& 0.396038 -7.77987\times10^{-5}X_e \\ \nonumber
& &-3.03313\times10^{-8}X_e^2.
\end{eqnarray}

\begin{figure*}
\includegraphics[width=16.5cm]{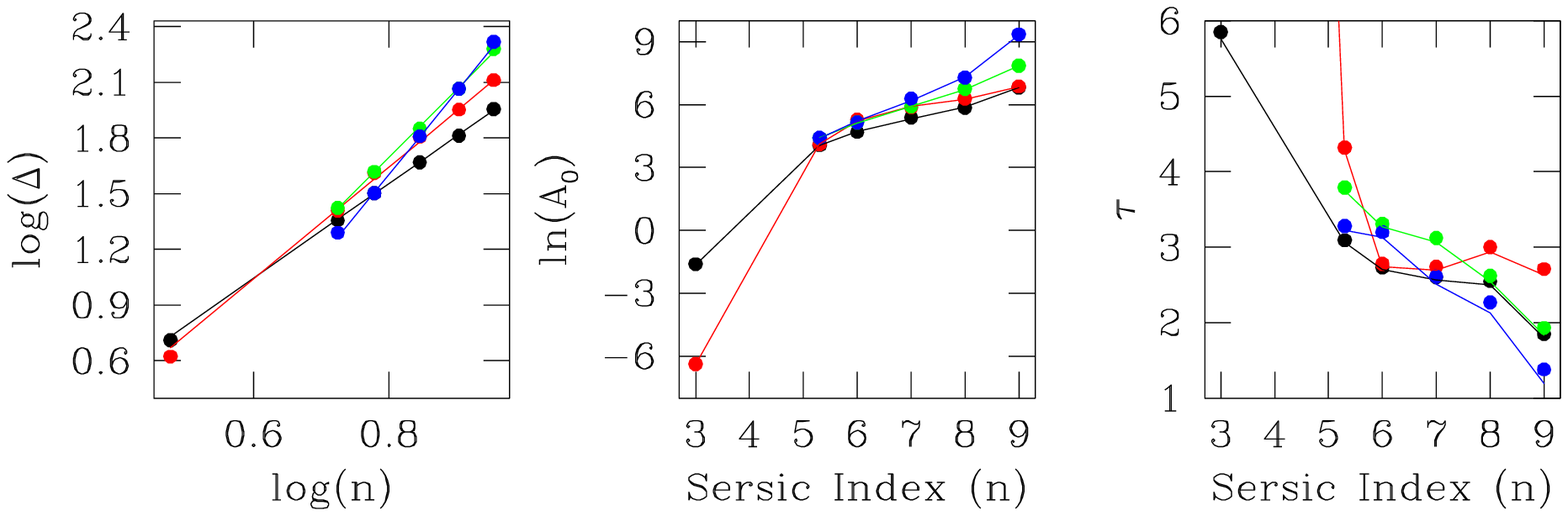}
\caption{\label{fig:abtau} Trends in the three different parameters
that define the exponential fall-off of $\Upsilon_{out}$ with
$\Upsilon_{in}$, with changes in $n$ and $X_e$; marked in filled
circles of colours that are distinctive of the $X_e$ value used -
black for $X_e$=726 pc, red for $X_e$=1000 pc, green for $X_e$=1400 pc
and blue for $X_e$=2000 pc. These trends have been spotted during the
parametric analysis that we undertake with our model galaxies. The
recovered functional dependence that defines these trends are marked
in solid lines of colour corresponding to the value of $X_e$. These
functional dependences were recovered using model galaxies with
$\{n=3,5.3,6,7,8,9\:{\rm all X_e}\leq 2000{\rm pc}\}$}.
\end{figure*}

\subsection{Relation between $\Upsilon_{out}$ $\&$ $\Upsilon_{in}$}
\noindent
The data from the assorted models indicate that $\Upsilon_{out}$ is
noted to fall exponentially with $\Upsilon_{in}$; in fact, a good fit
to this relation is given by the following equation:
\begin{equation}
\Upsilon_{out} = \Delta(n, X_e) + A_0(n, X_e)\exp[-\Upsilon_{in}/\tau(n, X_e)],
\end{equation}
where $\Delta$, $A_0$ and $\tau$ are functions of $X_e$ and $n$. The
relations between $n$ and $\Delta$, $A_0$ and $\tau$ are shown in the left,
middle and right panels of Figure~\ref{fig:abtau}, for the four
different $X_e$ values that we use. From our models, we seek the
functional form of these relations; these recovered functions are
over-plotted in solid lines, on the data in the three panels in
Figure~\ref{fig:abtau}. This comparison indicates that the known
trends in $\log\Delta$, $\ln{A_0}$ and $\tau(n)$, with increasing $\log(n)$
and $n$, for the assorted $X_e$ values, are well replicated by the
predicted functional forms of the relevant quantities.

We advance the following formulae for $\Delta$, $A_0$ and $\tau$:
\begin{eqnarray}
\label{eqn:delta}
\log[\Delta(n, X_e)] &=& C_0(X_e) + C_1(X_e)\log(n) \:\textrm{where} \\ \nonumber
C_0(X_e) &=& 0.18832 -8.40018\times10^{-4}X_e- \\ \nonumber
& &1.19449\times10^{-7}X_e^2 \quad \textrm{and} \\ \nonumber
C_1(X_e) &=& 1.15563 + 0.00207X_e - \\ \nonumber
{} & & 2.06978\times10^{-7}X_e^2
\end{eqnarray}
\begin{eqnarray}
\label{eqn:A0}
\ln[A_0(n, X_e)] &=& D_0(X_e) + D_1(X_e)n + D_2(X_e)n^2 \\ \nonumber
& & +D_3(X_e)n^3 \quad \textrm{where} \\ \nonumber
D_0(X_e) &=& 480.078 -1.32685X_e \\ \nonumber
& & + 0.00106312X_e^2 \\ \nonumber
& & -2.63527\times10^{-7}X_e^3, \\ \nonumber
D_1(X_e) &=& -192.617 +0.537814X_e \\ \nonumber
& & -0.000431932X_e^2 \\ \nonumber
& & +1.07350\times10^{-7}X_e^3, \\ \nonumber
D_2(X_e) &=& 24.9361 -0.0699530X_e \\ \nonumber
& & +5.63996\times10^{-5}X_e^2 \\ \nonumber
& & -1.40766\times10^{-8}X_e^3 \quad \textrm{and} \\ \nonumber
D_3(X_e) &=& -1.03518 +0.00293420X_e -\\ \nonumber
& & 2.37672\times10^{-6}X_e^2 \\ \nonumber
& & +5.96666\times10^{-10}X_e^3.
\end{eqnarray}
\begin{eqnarray}
\label{eqn:tau}
\tau(n, X_e) &=& \tau_0(X_e) + \tau_1(X_e)n + \tau_2(X_e)n^2 +\\ \nonumber 
& &\tau_3(X_e)n^3 +\tau_4(X_e)n^4 \quad \textrm{where} \\ \nonumber
\tau_0(X_e) &=& -3650.81 +8.99277X_e -0.00649587X_e^2 \\ \nonumber
& & +1.43425\times10^{-6}X_e^3, \\ \nonumber
\tau_1(X_e) &=& 1903.74 -4.64150X_e +0.00331538X_e^2 \\ \nonumber
& & -7.22203\times10^{-7}X_e^3, \\ \nonumber
\tau_3(X_e) &=& -367.254 +0.885327X_e -0.000623864X_e^2 \\ \nonumber
& & +1.33667\times10^{-7}X_e^3 \quad \textrm{and} \\ \nonumber
\tau_4(X_e) &=& 31.1895 -0.0742165X_e +5.14858\times10^{-5}X_e^2 \\ \nonumber
& & -1.08171\times10^{-8}X_e^3.
\end{eqnarray}

These relations are valid for all model systems with $n\geq$3 and that
too, for $n$=3, only systems with $X_e\leq$4000pc and for $n$=4,
$X_e\leq$5000pc.

\subsection{Photometry in other wave-bands}
\label{sec:photo}
\noindent
It is always possible that the observed photometry is presented in a
wave-band different from the ACS $z$-band for which we predict the
above relations between the properties of the raw $M/L$, though the
$z$-band is in general a better choice for high redshift systems than
a bluer waveband. If the available photometry is in a band different
from $z$, (say the $w$-band), then the factor by which every value of
$\Upsilon_{in}$ and $\Upsilon_{out}$ should be changed, is obtained
from the following considerations. We realise that if $w$ is such that
a model galaxy is brighter in $z$ than in the $w$-band, then the $M/L$
ratio by which the inner and outer parts of the brightness profile in
$z$ need to be scaled, are smaller than the same by which the profile
in $w$ needs to be scaled. In fact, the value of $\Upsilon_{out}$
corresponding to observations in $w$ should be scaled by a factor of
$10^{0.4(L_{\odot w} - L_{\odot z})}$. Here $L_{\odot w}$ is the solar
absolute magnitude in the $w$-band. In the ACS $z$-band $L_{\odot
z}$=4.52. However, this transformation into the $w$-band is really an
approximation since we assume all the way that $X_e$ is the same over
wavebands. This is not true \citep{temi_08, ko_05}.

\subsection{Distance to the galaxy}
\label{sec:D}
\noindent
It is important to enquire about the stability of the posited forms of
$\Upsilon_{out}$ and $\Upsilon_{in}$ when the distance to a galaxy is
different from what has been used in the models, namely 17 Mpc
(approximate distance to Virgo). The only influence of the distance
$D$ (in Mpc) to the observed galaxy is in affecting the luminosity
density distribution that is obtained be deprojecting the observed
surface brightness profile, through a term that is linear in
$D^{-1}$. It is this luminosity density profile that is scaled by
$\Upsilon_{in}$ and $\Upsilon_{out}$ in the inner and outer parts of
the galaxy. Thus, if the galaxy is in reality further than 17 Mpc, the
implementation of the suggested $M/L$ values would amount to an
overestimation of the luminosity density. To compensate for this, we
need to modulate the relevant $\Upsilon_{out}$ and $\Upsilon_{in}$ values 
by the factor $D/17$.

\section{Tests}
\noindent
In this section we discuss the testing of our advanced parametric
forms of $\Upsilon_{out}=f(n, X_e,\Upsilon_{in})$ (Equation 9 with
inputs from 10, 11 and 12) and $\Upsilon_{in}^{max}=g(n, X_e, \alpha)$
(Equation 8). The scheme is tested on models 
that were excluded from the fitting exercise that resulted in our 
identification of the forms of the functions $f$ and $g$.

\begin{figure}
\includegraphics[width=7.5cm]{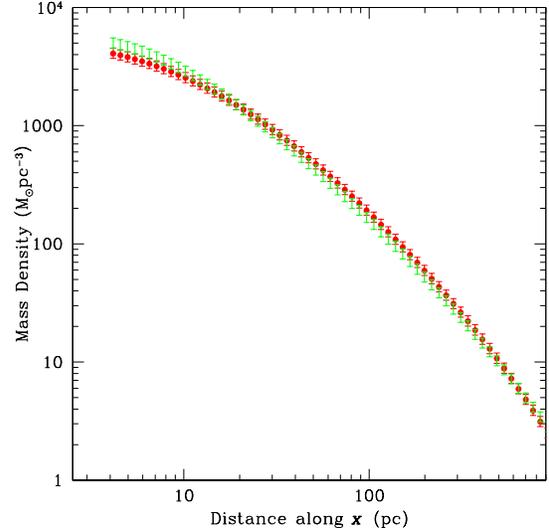}
\caption{\label{fig:compare} Figure to compare the predicted total mass density
distribution with $x$ (in red) with the known (model) mass density
profile (in green), for the model described by $n$=3.5 and $X_e$=3000
pc. For an $\alpha$ of 5, this model yields
$\Upsilon_{in}^{max}\approx$5.8 and $\Upsilon_{out}\approx$0.5. The
profile in red results on scaling the luminosity profile of this model
galaxy by the raw $M/L$ distribution that is defined by these
``inner'' and ``outer'' amplitudes and subsequently smoothing the
discontinuous mass density profile thus obtained, (using the smoothing
prescription mentioned in the text).}
\end{figure}

Figure~\ref{fig:compare} shows a comparison of the predicted and model
mass density profiles along the $\bf{x}$-axis for the model galaxy
that has a Sersic index of 3.5 and $X_e$=3000 pc. $\alpha$ is set to 5
$\Longrightarrow\sigma_0\approx$250 kms$^{-1}$, averaged over a radius of about
$X_e$/8=375 pc. For this system, we extract
$\Upsilon_{in}^{max}\approx$5.8 and $\Upsilon_{out}\approx$0.5. This
pair of $\Upsilon_{out}-\Upsilon_{in}$ values are used to scale the
luminosity density distribution for this model galaxy which is then
smoothed to offer a final mass density distribution along the major
axis. These $\Upsilon_{out}-\Upsilon_{in}$ values imply a predicted
mass density distribution (in red in Figure~\ref{fig:compare}) that
tallies favourably with the model (in green in
Figure~\ref{fig:compare}).

\begin{figure}
\includegraphics[width=7.9cm]{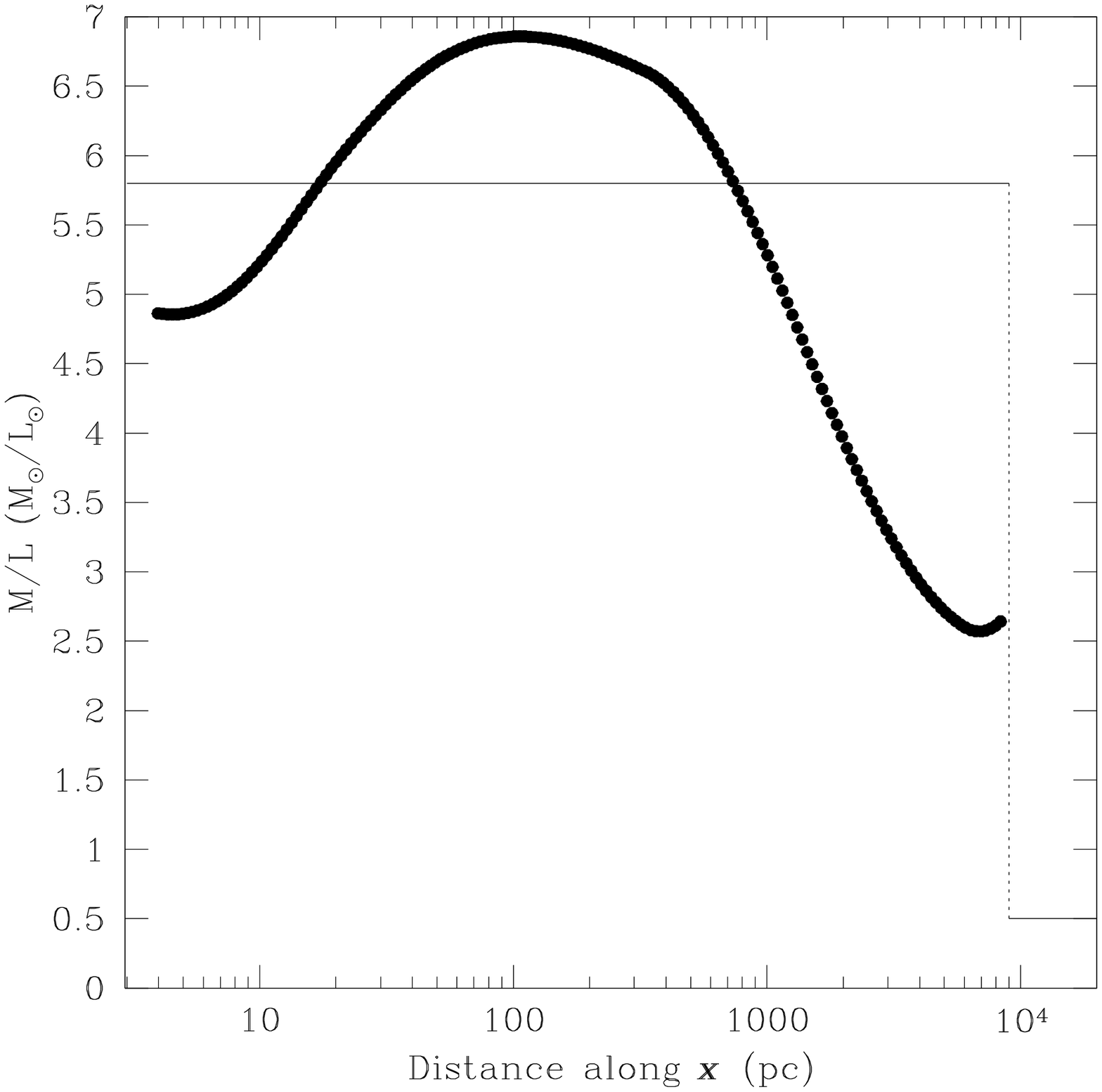}
\caption{\label{fig:compare_ml} Figure to compare the final total
(local) $M/L$ distribution with the initially chosen raw $M/L$ profile
(in broken lines), for the test galaxy the total mass of which is
presented in Figure~\ref{fig:compare}. The final $M/L$ is presented
within 3$X_e$. The raw $M/L$ profile is the same as that shown in
Figure~\ref{fig:scheme}; as is apparent from this figure, it is
significantly different from the final $M/L$ profile. }
\end{figure}

The predicted mass density distribution when compared to the luminosity
density distribution offers the final or smoothed $M/L$ which is shown
in Figure~\ref{fig:compare_ml} to be significantly different from the raw
$M/L$ profile (in this black line).

\begin{figure*}
\centering{
  $\begin{array}{c c}
  \includegraphics[height=.3\textheight]{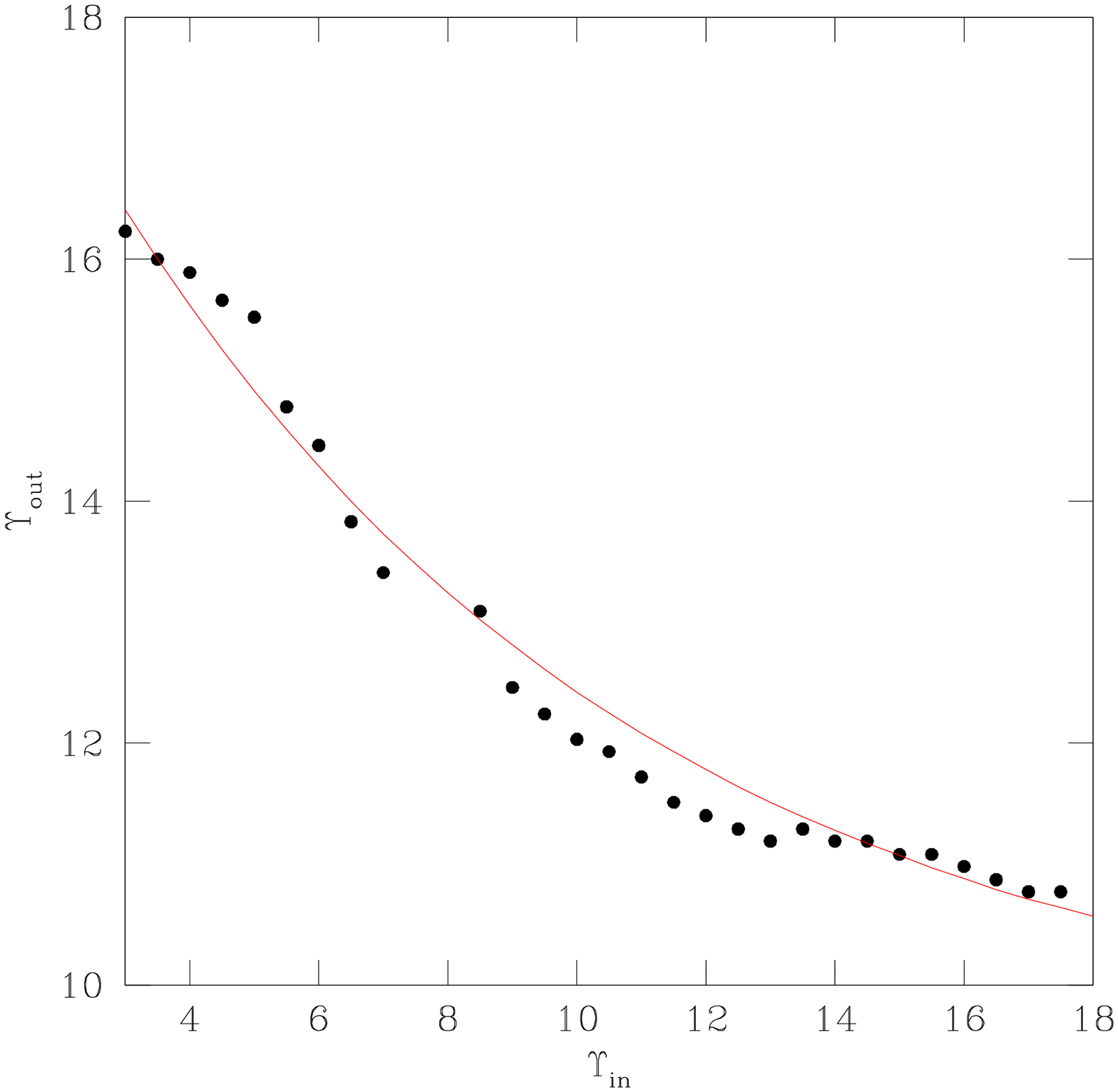} &
  \includegraphics[height=.3\textheight]{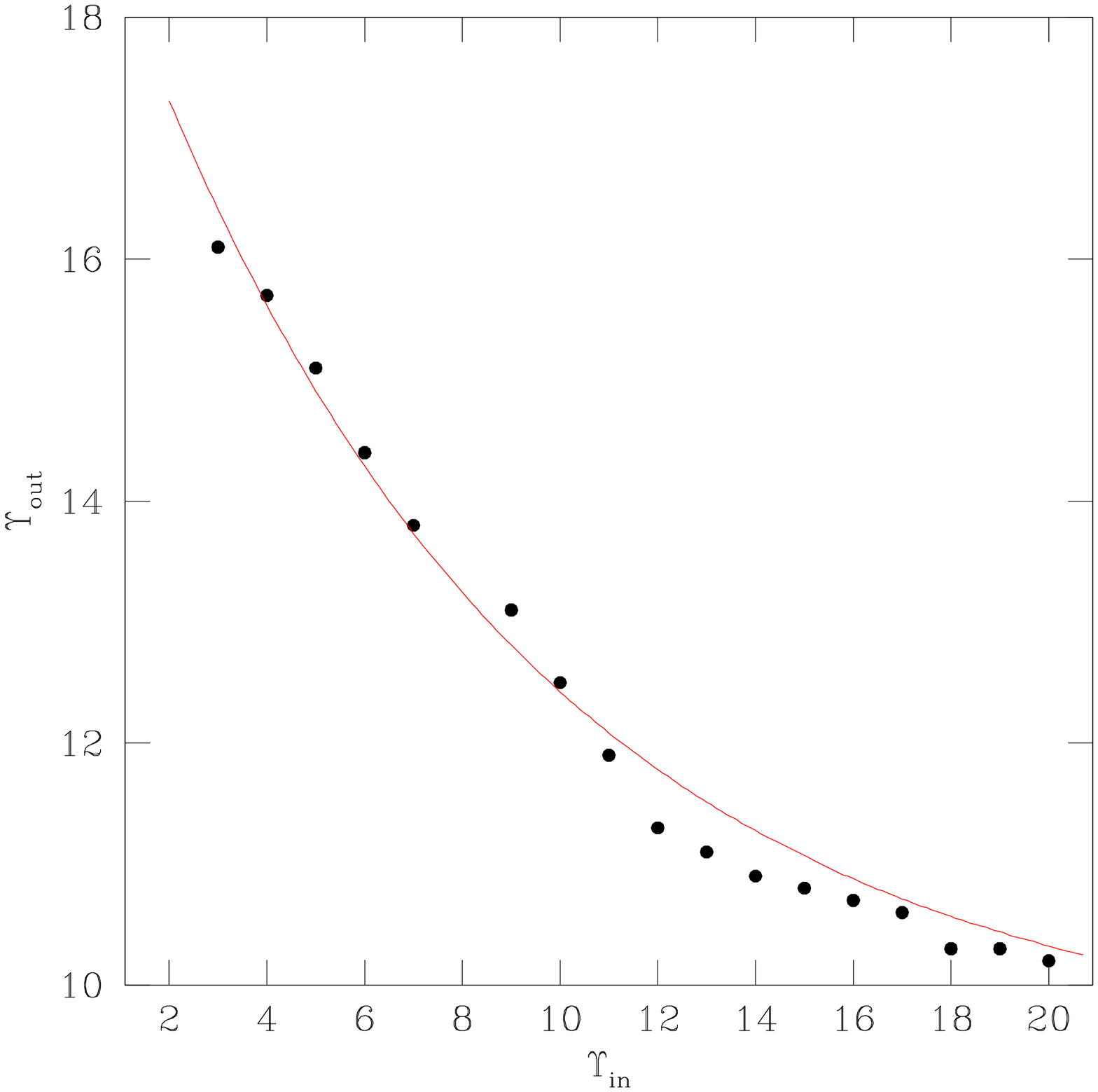} \end{array}$
\caption{\label{fig:validate} Comparison of model $\Upsilon_{out}$, as obtained from our calculations (black filed
circles) and predicted $\Upsilon_{out}$ (red line), for given
$\Upsilon_{in}$, corresponding to model systems with Sersic index = 4
and $X_e$=726 pc (left) and $X_e$=1400 pc (right). The predicted curves are
obtained by using the values of $n$ and $X_e$ for the model test
system, in Equations 9, 10, 11 $\&$ 12. The calculated
$\Upsilon_{out}$ values result from our analysis.}}
\end{figure*}

However, this was the example of one given model galaxy, characterised
by a given central velocity dispersion. Tests were undertaken to
validate the relations predicted between the properties of the raw
$M/L$ and the photometric parameters of model galaxies, across the
full range of $\sigma_0$. The tests were carried out with model
galaxies with Sersic index of 4 and $X_e$ of 726 pc and 1400 pc. In
Figure~\ref{fig:validate}, we see the calculated $\Upsilon_{out}$
value corresponding to a given $\Upsilon_{in}$, for which
compatibility is noted between the known (model) and predicted total
mass density distributions, to 3$X_e$. Such $\Upsilon_{out}$ values
are shown in the two panels of this figure, in black dots. 

It is to be noted that these calculated values of $\Upsilon_{out}$
very closely straddle (within errors of $\pm$5$\%$) the analytical
relationship between $\Upsilon_{out}$ and $\Upsilon_{in}$ that is
predicted in Equations~9, 10, 11 and 12, for given $X_e$ and $n$; in
this case for $n$=4, $X_e$=726 pc (left) and $n$=4, $X_e$=1400 pc
(right). Thus, these tests offer confidence in the formalism that we
suggest.

\section{Applications}
\noindent
In this section, we check out the efficacy of the advanced scheme in
recovering gravitational mass density distribution of two elliptical
galaxies, NGC~4494 and NGC~3379, to 3$X_e$. Our predicted mass
distributions are compared to independent dynamical mass models for
these systems.

\subsection{NGC~3379}
\noindent
NGC~3379 was reported by \cite{roma_science} and \cite{douglas_07} to
contain very little dark matter on the basis of a Jeans equation
analysis of the kinematics of around 200 planetary nebulae (PNe) that
reside in the dark halo of this galaxy. An independent estimation of
the distribution of the total mass density of this galaxy was
performed by Chakrabarty, 2009 (submitted to {\it AJ}), by
implementing these PNe velocities in the Bayesian algorithm CHASSIS
\citep{saha_dal, dal_zwart}. As acknowledged by Chakrabarty (2009),
these mass distributions from CHASSIS indicate somewhat higher masses
than the estimates of \cite{douglas_07}, owing to the
assumption of isotropy within CHASSIS. While details of such mass
estimation techniques are irrelevant to the current work, here we
present a comparison between the mass density profile obtained from
our formalism with the same obtained from CHASSIS. We also present a
comparison between the cumulative mass result $M(r)$ via the quantity
defined as $v_c = \sqrt{GM(r)/r}$, where $v_c$ is referred to as the
circular velocity.

\begin{figure*}
\includegraphics[width=13.9cm]{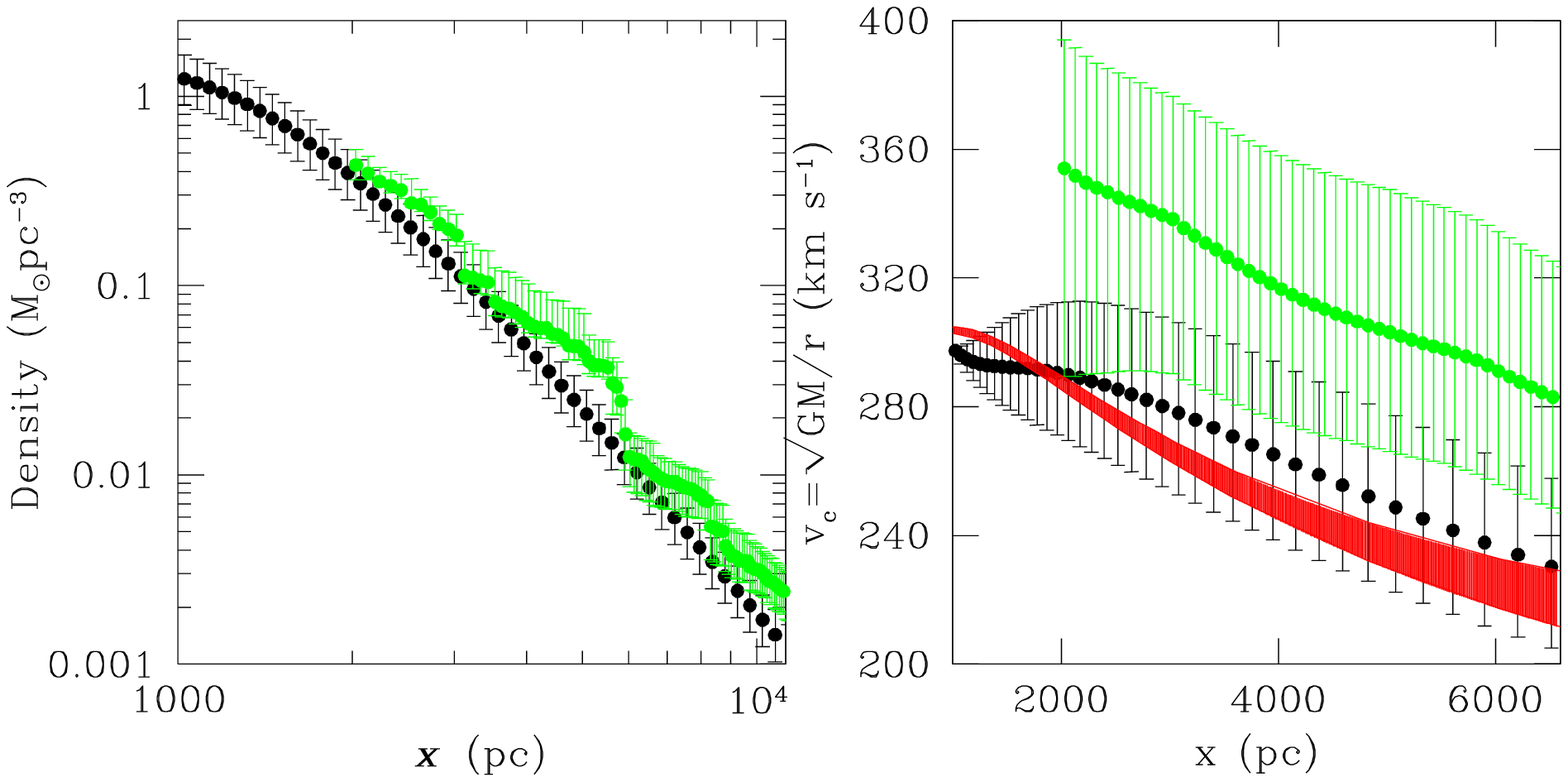}
\caption{\label{fig:3379} The radial distribution of the recovered
total mass density is depicted in black, on the left. On this is
superimposed the mass density profile obtained by implementing the
kinematic information of a sample of planetary nebulae in NGC~3379
(the same sample as used by Douglas et. al, 2007), in the Bayesian
algorithm CHASSIS (Chakrabarty 2009, communicated to {\it AJ}). The
quantity $v_c$, as recovered from these mass distributions are shown
in corresponding colours on the right, with the $v_c$ profile from
\cite{douglas_07} overplotted in red.}
\end{figure*}

Our predictions are based on the analysis of the photometry of
NGC~3379 in the $B$-band \citep{cappacicoli_90} and the data for
projected central velocity dispersion, as given by
\cite{statler_99}. Using this $\sigma_p$ data along the major axis of
the galaxy, \citep[Table~1A of][]{statler_99}, we get an $\alpha$ of
about 7.5. The photometry suggests an $X_e$ of 2.2 kpc. Additionally,
we deproject the $B$-band surface brightness profile using the
Bayesian deprojection algorithm DOPING \citep{doping}. We consider
NGC~3379 to be at a distance of 11 kpc, as in \cite{douglas_07}. Our
predicted values of (upper limit on) $\Upsilon_{in}$ and
$\Upsilon_{out}$, as modulated by differences in the wave-band of the
available photometry and distance to the system, are about 9.3 and
5.9.

The resulting mass density distribution that we advance for NGC~3379
within 3$X_e$, is depicted in black on the left of
Figure~\ref{fig:3379}. This is compared to the mass model identified by
CHASSIS. $v_c$ derived
from our calculated mass distribution, from the dynamical mass
modelling by CHASSIS and that reported by \cite{douglas_07} are
represented in the right panel of Figure~\ref{fig:3379}.

The comparison of the total mass density profile indicates the clear
trend for our predicted mass distribution to be on the lower side
compared to the gravitational matter density provided by CHASSIS. This
is only to be expected since CHASSIS in its current form, assumes
isotropy, which affects CHASSIS idiosyncratically to spuriously
enhance mass density, as acknowledged by Chakrabarty
(2009). Consequently, the $v_c$ profile advanced by our work is also
on the lower side of the profile that follows from CHASSIS, though our
result compares better with the result advanced by \cite{douglas_07}.

\begin{figure*}
\includegraphics[width=13.9cm]{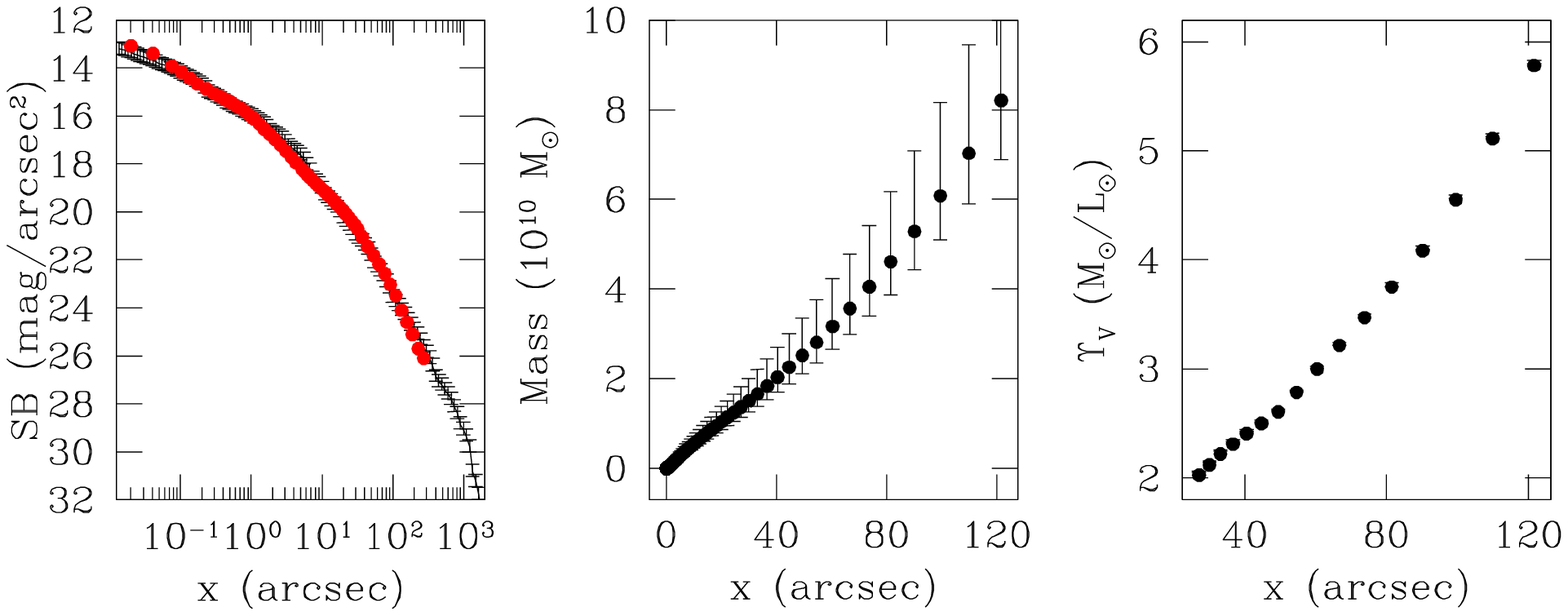}
\caption{\label{fig:4494} Figure to bring out the results for the
galaxy NGC~4494. The left panel depicts a projection of the luminosity
density distribution that we estimate for the galaxy (in black),
compared to the $V$-band surface brightness data (in red). The total
gravitational mass distribution of the galaxy is represented in the
middle panel while the mass-to-light ratio on the $V$-band is shown on
the right.}
\end{figure*}

\subsection{NGC~4494}
\noindent
NGC~4494 is a nearby elliptical, the mass distribution of which to 7
effective radii has recently been presented by \cite{napo_08}. The
distance to this galaxy is given as 15.8 Mpc by \cite{napo_08}. The
$V$-band photometry of this galaxy is presented in Table~A1 of
\cite{napo_08}. This surface brightness profile indicates an $X_e$ of
3.37 kpc and is deprojected, given the radial variation of the
projected eccentricity (assuming an oblate geometry and edge-on
viewing). The projection of such recovered luminosity
density distribution is shown in the left panel of
Figure~\ref{fig:4494}, compared to the surface brightness data of this
galaxy (in red). \cite{napo_08} also cite the central velocity
dispersion of this system as 150 kms$^{-1}$ \citep{paturel_03}.
Considering this $\sigma_p$ to be the dispersion averaged over
$R_e/8$, where $R_e$=48''.2 for this galaxy, we obtain an $\alpha$ of
2. Our predicted values of $\Upsilon_{in}$ and $\Upsilon_{out}$ are
about 2 and 0.121. These are used to characterise the raw $M/L$, which
when smoothed, gives rise to the total mass distribution that is shown
in the middle panel of Figure~\ref{fig:4494}. 

The total $M/L$ ratio of NGC~4494, in the $V$-band is shown on the
right panel of Figure~\ref{fig:4494} while the enclosed gravitational
mass distribution is depicted in the middle panel. This is very
similar to the radial mass distribution of NGC 4494, within the inner
3$X_e$, as reported by \cite{napo_08} (their Figure~13).

\section{Discussions and Summary}
\noindent
We have presented a novel mechanism for estimating total mass density
profiles of elliptical galaxies that can be described by a Sersic-type
surface brightness distribution. This formalism uses nothing in excess
of what is typically available in the observational domain, namely,
photometry and the central velocity dispersion profile. This allows
the implementation of this scheme even for systems at high
redshifts. (Such an implementation of this scheme will be presented in
a future contribution - Chakrabarty $\&$ Conselice, under
preparation). The advanced scheme uses the methodology presented in
Paper~I, in the context of Sersic galaxies.

To begin with, we generate a sample of luminosity density
distributions that project to Sersic brightness profiles, of assorted
values of $n$ and $X_e$. These toy galaxies are then assigned various
values of the local central mass-to-light ratio (parametrised by
$\alpha$). For each such configuration, we monitor the allowed range
of $\Upsilon_{in}$ and the value of $\Upsilon_{out}$ that corresponds
to any such $\Upsilon_{in}$. This pair of
$\Upsilon_{in}-\Upsilon_{out}$ values define the raw or unsmoothed
two-stepped local mass-to-light ratio distribution with $x$. The local
mass density distribution in the system (to 3$X_e$) is judged by
scaling the luminosity density profile by the raw $M/L$ profile and
then smoothing the result by the prescribed smoothing routine (two
successive applications of a box filter of size corresponding to
$X_e$).

The dependence of $\Upsilon_{out}$ on $\Upsilon_{in}$ is found to be
well approximated with an exponential decline that is described by
three free parameters, which are actually all functions of the model
characteristics, namely, $n$ and $X_e$. The dependence of these three
parameters on $n$ and $X_e$ are extracted from our model-based study
and analytical fits to these trends are presented. Such fits are noted
to be favourably represented by polynomials.

The reliance of the band of the allowed values of $\Upsilon_{in}$ on
$\alpha$ is also quantified as ranging from $\alpha$ to
$\Upsilon_{in}^{max}$, where the form of the function
$\Upsilon_{in}^{max}(n, X_e, \alpha)$ is also found to be polynomial
in nature.

\subsection{Deviation from predicted trends?}
\noindent
Even though the polynomial fits are found to do a good job for most
configurations, for certain models, the predictions appear incorrect
(by no more than 10$\%$), eg. $\tau$ may be judged under-estimated for
$n$=8 and 9 for the configuration: $X_e$=2000 pc. However, this should
be interpreted as erroneous recording of $\Upsilon_{out}$ for these
$\Upsilon_{in}$ values from our models, rather than a failure of our
predictions.  Typically, such errors emanate from lack of refinement
in the step-sizes used in our work and is potentially amendable.

\subsection{Applicable to$\ldots$}
\noindent
As delineated in Section~\ref{sec:models}, the formalism that we
advance here, is applicable only to galaxies with Sersic indices $n
\geq 3$. It is only for such systems that the luminosity density
profile is steep enough, i.e. falls quickly enough to ensure that in
the ``outer'' parts of the system the luminous matter density is
completely overwhelmed by the dark matter density, so that the total
matter density is effectively a representation of the dark matter
distribution. Thus, changes in the central $M/L$ or $\alpha$ do not
affect the $M/L$ in these ``outer'' parts. Now, to qualify ``outer'',
we state that this region corresponds to $x\sim 3X_e$. This is anyway
true, only if the luminous matter fraction is not too small, in which
case, we need to settle for progressively lower $M/L$ in the outer
parts, as $\alpha$ increases.

\subsubsection{Effect of lowering Sersic Index}
\noindent
If however, the luminous density profile is too flat, (as for $n <$
3), the contribution of the luminous matter, to the total mass, is
significant, even in ``outer'' parts. In order to achieve similarly
shaped total mass profiles, the value of $\Upsilon_{out}$ will need to
increase with increasing $\alpha$. Then, the exponential fall-off of
$\Upsilon_{out}$ with $\Upsilon_{in}$ will not be noted. Qualitatively
speaking, this is the reason why our formalism will not be valid for
galaxies with Sersic indices less than 3.

\subsubsection{Effect of increasing size}
\noindent
However, this increased flattening of the luminosity density profile
can be attained even for $n\geq$3, for large $X_e$ values. Thus, in
such systems, for the same reason as in the last paragraph,
$\Upsilon_{out}$ will increase with an increase in the inner $M/L$
instead of falling exponentially with increasing $\Upsilon_{in}$.
Thus, based on these considerations, on the basis of our experiments, we
find that the following galaxies {\it cannot} be accommodated within our
suggested formalism:
\begin{itemize}
\item $n$=3 and $X_e>$4000 pc,
\item $n$=4 and $X_e>$5000 pc,
\item $n <$3.
\end{itemize}

\subsubsection{What the ranges mean}
\noindent
However, it is important to keep in mind that these ranges have been
ascertained on the basis of a discretised scanning of the parameters
$n$ and $X_e$, of our two-parameter galaxy models. Thus, when we say
that models with $n<$3 cannot be explained with the predicted
relationships between $\alpha,\:,\Upsilon_{in},\:\Upsilon_{out}$, what
we really mean is that our experiments with a model with $n$=2
indicated failure of adherence to these predicted trends. It is very
much possible that a model with an intermediate Sersic index can be
explained by our predictions, such as a galaxy with $n$=2.3, say. In
other words, the ranges of $n$ and $X_e$ described above owes, to our
dealing only with models ascribed the aforementioned discrete values
(values mentioned in Section~\ref{sec:models}). Consequently, the
ranges listed above are a conservative interpretation.

The practical question to ask is, are these suggested ranges
compatible with real galaxies? As itemised in
Section~\ref{sec:models}, our choice is supported by the trends
observed for the Virgo cluster, within the ACS Virgo Cluster Survey,
in alliance with the predicted evolution towards greater
compactness in high-redshift systems \citep{ferreras_09}. Furthermore,
keeping in mind that a waveband dependence of effective radius exists,
our model $X_e$ values in the $z$-band fall within limits indicated by
observations \citep{laura_acs, nacho_01, barbera_05}.

Additionally, we question the relevance of the upper limits on the
effective radii at $n$=3 and 4, which are smaller than the range
permitted for higher values of $n$. As far as the correlation between
the shape parameter $n$ and log of effective radius is concerned, it
is well known that in general, half-light radius decreases with
decreasing $n$ \citep{boselli_08, pat_08, nacho_06, brown_03,
nacho_01}. In fact, such a correlation is physical and is hinted at by
the global relations between structural parameters; it is not merely a
reflection of the fitting procedure \citep{nacho_01}.

\subsubsection{Effect of model dark matter distribution}
\noindent
Of course, it could be perceived that such fate of the mass
distribution at $x\sim 3X_e$ would be dictated by the exact details of
the underlying dark matter distribution that defines the model
galaxy. Indeed, for the high values of $n$, the $M/L$ at $x \geq
3X_e$ ($\Upsilon_{out}$) is dictated by the details of the dark matter
distribution, but it is the {\it smoothing of this raw $M/L$ profile}
that gives us the end product, namely the total local $M/L$ ratio
profile of the galaxy upto the benchmark $x$ of $3X_e$. It is
precisely this smoothing procedure that brings into the final $M/L$
structure, elements of the mass distribution in the inner $3X_e$ in
the galaxy, as well as a relatively less significant signature of the
same that lies {\it outside the radius upto which the distribution is
sought}. The difference that smoothing makes to the raw $M/L$ profile,
as distinguished from the final $M/L$ profile, is clear in
Figure~\ref{fig:compare_ml}.

Thus, given such (1) implementation of our smoothing procedure and
that (2) the jump radius in the raw $M/L$ profile is itself $3X_e$, we
find that our experiments bear the fact that changing the mass scale
and length scale of the used NFW dark matter density distribution
across the wide ranges do not affect the mass
configuration for $x\lesssim 3X_e$, though further out in the galaxy, the
influence of changing the DM density distribution picks up
quickly. The consistency in the $\Upsilon_{out}-\Upsilon_{in}$
relation, noted with changes in $M_s$ and $r_s$ is brought out in the
left panel of Figure~\ref{fig:DM}. The allowed ranges in $M_s$ and
$r_s$ are provided in the right panel of Figure~\ref{fig:DM}. 

\subsubsection{Why choose NFW?}
\noindent
The modelling of the dark matter distribution in early type galaxies
is obviously indirect and consequently difficult and
unreliable. \cite{ferreras_08} suggest a large scatter in the outer
slope of the dark matter distributions that they recover for their
targeted galaxies. \cite{gavazzi_07} corroborate an NFW modelling of
the dark haloes of their sample galaxies, suggesting that the total
matter density tends to an overall isothermal form. Given this degree
of uncertainty, it might be argued that there is not much sense in
splitting hair to decide between an NFW and isothermal dark matter
distributions, as long as the dark haloes that we probe in our analysis
are compatible with observations or simulations. 

\begin{figure*}
\centering{
  $\begin{array}{c c}
  \includegraphics[height=.3\textheight]{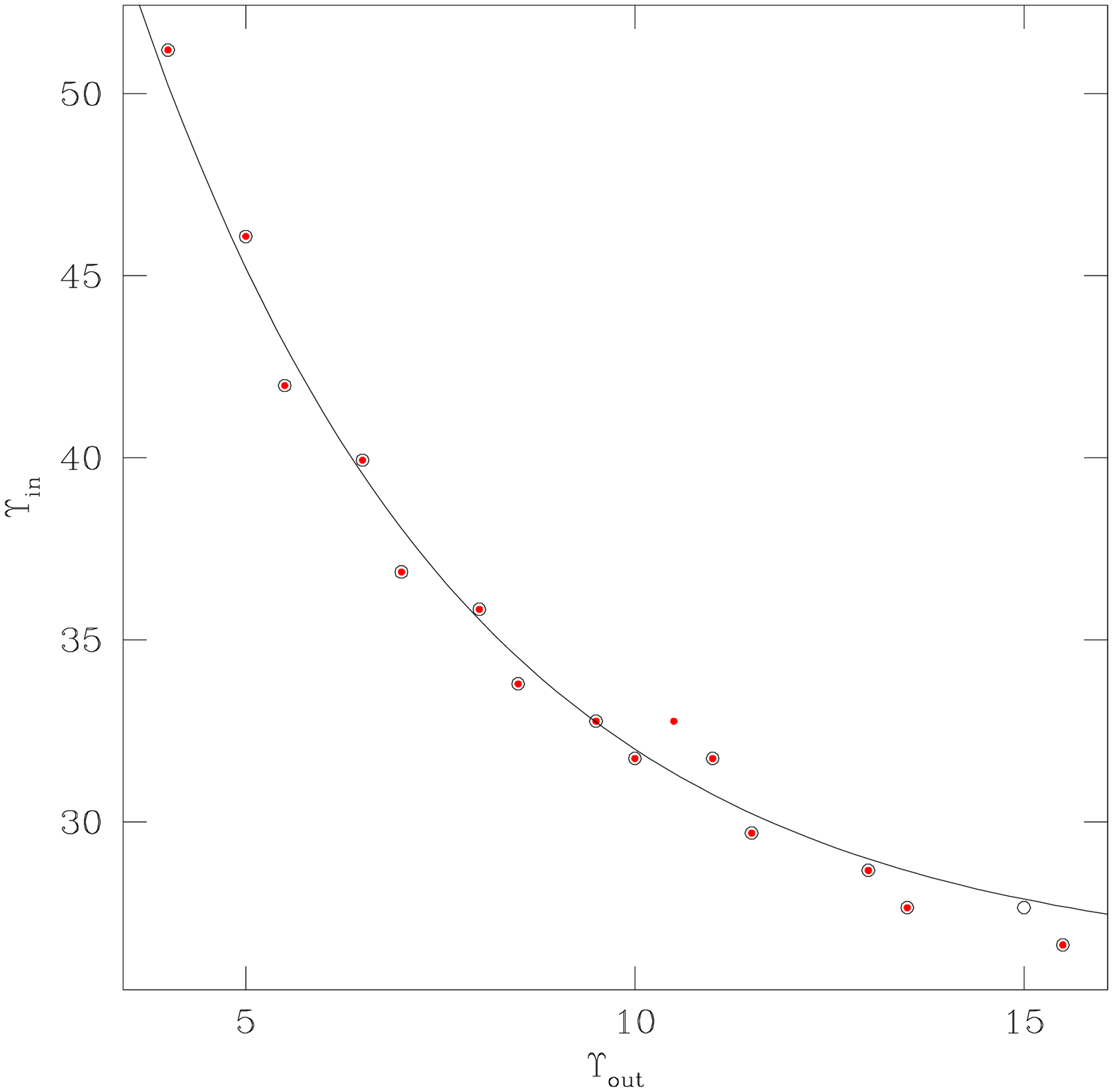} &
  \includegraphics[height=.3\textheight]{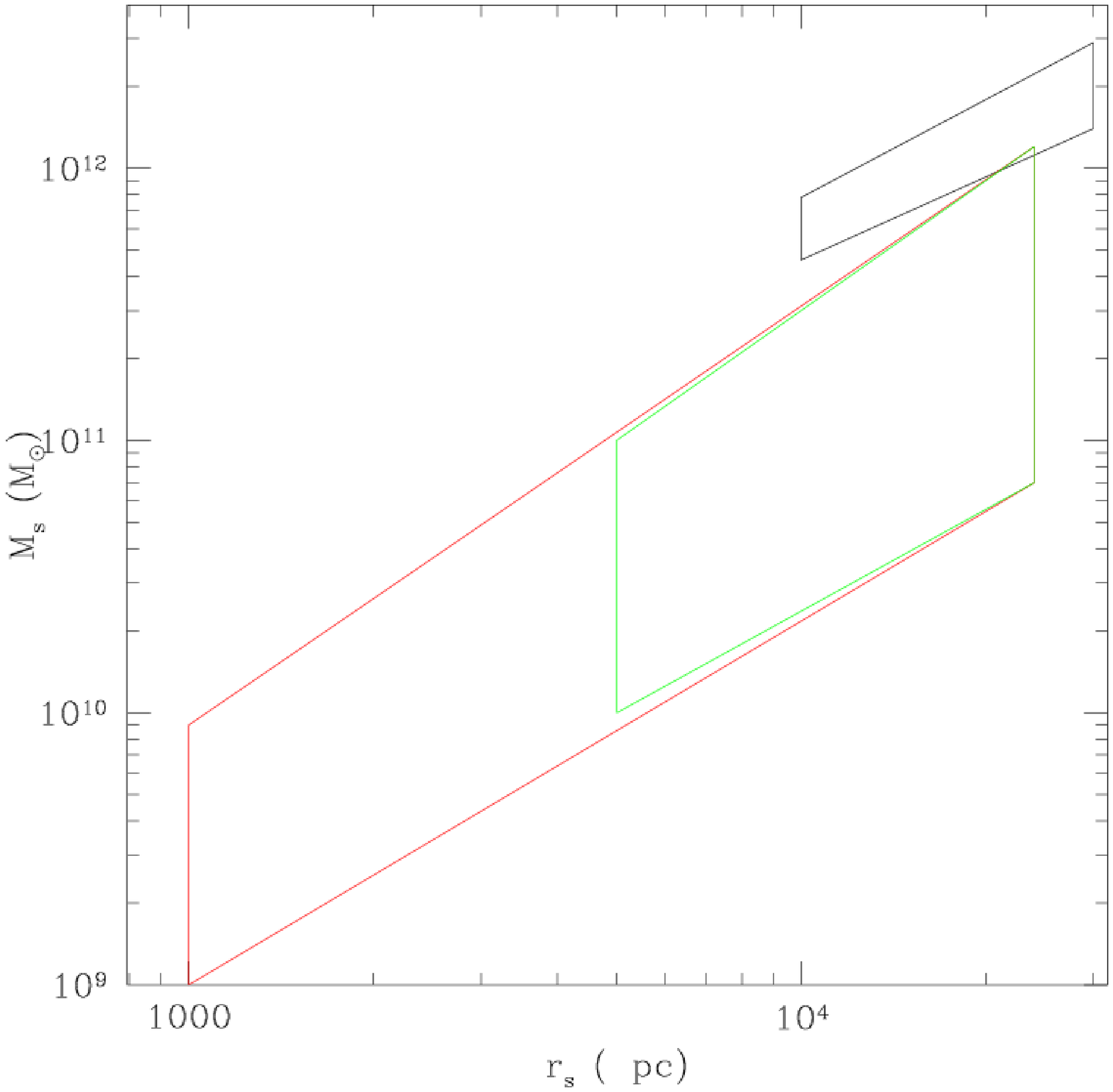} \end{array}$
\caption{\label{fig:DM} The left panel depicts the relation between
$\Upsilon_{out}$ and $\Upsilon_{in}$, obtained using the model with
$n$=5.3 and $X_e$=1000pc, from two different runs that seek this
relation, given NFW haloes characterised by $(M_s, r_s)$ values within
the quadrilaterals in black (see right panel) and in red (see right
panel). When the halo density parameters are as contained within the
black quadrilateral, the obtained $\Upsilon_{out}-\Upsilon_{in}$
relation is shown in open black circles in the left panel. When the
halo density is characterised by $M_s$ and $r_s$ values as points
inside the red quadrilateral on the right, the
$\Upsilon_{out}-\Upsilon_{in}$ relation is shown in red filled circles
in the left panel. The predicted $\Upsilon_{out}-\Upsilon_{in}$
relation in shown on the left in solid black lines. The green
quadrilateral represents the subset in the $M_s-r_s$ space, which
defines the dark matter density distributions for which concurrent $f$
and $g$ functions were recovered in general runs done with assorted
models, within our scheme.}}
\end{figure*}

This is indeed the case, when we compare our ranges of halo
characteristics (right panel of Figure~\ref{fig:DM}) to the suggestion
by \cite{combo_06} that the recovered range of virial masses of their
NFW model to be $[3.9\times10^{11}, 7.1\times 10^{11}]$
M$_\odot$. Such a mass range is compatible with the suggestion by
\cite{hoekstra_06} for a fiduciary galaxy of luminosity $L_B=10^{10}$
L$_{B\odot}$ that has an NFW profile \citep[$M_{200}\in {[10, 6.4]}\times
10^{11}$M$_\odot$, as indicated by Figure~4 in][]{hoekstra_06}.
This mass range corresponds to a range of about 8 kpc to about 30 kpc
for $r_s$. In fact, we cover these ranges in the runs, the results of
which are presented in Figure~\ref{fig:DM}, and also scan haloes of
lower masses.  For various haloes defined within the red and black
quadrilaterals in Figure~\ref{fig:DM}, compatible
$\Upsilon_{out}-\Upsilon_{in}$ relations were recovered.

This explains our choice of sticking with the NFW prescription for our
dark halo model. For reasons similar to what we explain above, we
suggest that upon smoothing, haloes of varying shapes but similar
masses within 3$X_e$ are not expected to affect the mass distribution
within 3$X_e$ in the massive ellipticals that we deal with in this
formalism.

Thus, we see that our methodology is valid, irrespective of selecting
haloes defined by widely different $M_s$ and $r_s$ values, (i.e. NFW
haloes defined by points inside the black and red quadrilaterals in
Figure~\ref{fig:DM}). This motivates us to choose to work with a
smaller subset of all the haloes explored in these runs; in particular,
we choose to work with haloes defined by the more massive half of the
$\log(M_s)$ range that limits the red quadrilateral. This chosen
subset in the $M_s-r_s$ space is bound by the green quadrilateral in
Figure~\ref{fig:DM}.

\subsubsection{Effect of model geometry}
\noindent
In addition to details of the DM distribution, we admit that the model
structure does include the free parameters that describe the Sersic
model of density distributions in general triaxial galaxies, as it
ideally should, though this work has full potential of being extended
into two-dimensions and include an axisymmetric description of the
galaxy. Now, the triaxiality of the models needs to be defined in
terms of a chosen geometry and intrinsic eccentricities - it is of
course not possible to constrain such characteristics from
observations alone.

In our models, these were fixed as oblateness with an axial ratio of
0.7. It is indeed possible that the unknown functions $f$ and $g$ that
we attempt to constrain, harbour dependence on these intrinsic
geometric factors. However, the preparatory assumptions involved in
the deprojection of an observed brightness distribution, namely the
underpinning of intrinsic geometry and viewing angle, are essentially
unconstrained, unless the system is favourably inclined or
flattened. In other words, the unknown in our modelling are the usual
quantities that render deprojection non-unique.

As for the specification of the ellipticity in our models, a typical
value has been adopted - \cite{padilla_08} suggest an axial ratio
distribution for a large sample of SDSS ellipticals, with a mode in
the range of 0.6 to 0.8. Inspired by this, we use an axial ratio of
0.7 in our models. Again, the uncertainty in intrinsic ellipticity,
cannot be known for a general observed galaxy. The quantification of
the deprojection effects on the formulae provided above is possible,
at least in a statistical sense, and it is envisaged that the same
will be pursued in the future.

\subsection{What if central super massive black holes?}
\noindent
The case of a central mass condensation in the system was dealt with
in Paper~I, in reference to the example of the galaxy M87 - the
recovered mass distribution of M87 was demonstrated to be consistent
with the same obtained from kinematical considerations. If
independent measurements indicate an observed system to harbour a
massive central mass condensation, then the scheme delineated in
Paper~I will be followed.

\subsection{Why $X_e$ instead of $R_e$?}
\label{sec:X_e}
\noindent
The usage of $X_e$ instead of $R_e$ is preferred since $X_e$ as per
its definition here, as well as in Paper~I, is derived solely from the
inputs to the methodology, namely the surface brightness profile along
the semi-major axis which we consider to be along the ${\bf{\hat
x}}$-axis. Thus, any changes to the shape of the brightness profile
will be directly reflected in a linear change in $X_e$ but not
necessarily so in the half-light radius that is estimated from
isophotal analysis. In fact, here we use the same definition for $X_e$
as in Paper~I, (see Section~2).

The usage of $X_e$ should not be cause for concern since $X_e$ is
merely one definition of the semi-major axis effective
radius and reduces to the conventional definition of the major-axis
effective radius for $n$=4\footnotemark. 
When an unknown
galaxy is being analysed within this formalism, its half-light radius
is just as much an unknown as is the $X_e$ that we define here. Thus,
there is no loss of connection with observations by the implementation
of $X_e$. 

\footnotetext{Semi-major axis effective radius has been used before, for
example by \cite{nacho_06}}.

\subsection{Effect of choice of smoothing prescription}
\noindent
It merits mention that other choices for the smoothing
prescription and $x_{in}$ may also work, but here we concentrate on
the above mentioned configuration and the specification of
$\Upsilon_{in}$ and $\Upsilon_{out}$ are accordingly unique to these
choices.

\subsection{Effect of choice of estimation of central local $M/L$}
\noindent
In a similar context, it may be argued that the definition of $\alpha$
that we use herein will leave an imprint. We equate $\alpha$ to the
central local $M/L$, inspired by this result that is achieved in
PaperI. Here the total gravitational mass found enclosed within the
radius $x_0$ is $M(x_0)$ where $M(x_0)$ is linked to the 3-D velocity
dispersion at $x_0$ via: $\sigma_0^2 = GM(x_0)/x_0$. Then, according
to our definition, $\alpha$ is given as the ratio of $M(x_0)$ and the
enclosed light within $x_0$. For other definitions of $\alpha$,
other ranges of $\Upsilon_{in}$ will be valid (and therefore other
forms of dependences of $\Upsilon_{out}$ on $\Upsilon_{in}$, for the
same galaxy). Thus, the pairs of $f$ and $g$ functions that we advance
here, work for the used choice of the definition.

We would like to emphasise that the {\it extraction of the exact value of
the mass enclosed within $x_0$ (and therefore of $\alpha$), from a
measurement of $\sigma_0$, is not the point of this exercise};
uncertainties in this extraction do not undermine the advanced results
either, as long as the galaxy at hand is ``not too'' aspherical or
anisotropic at $x_0$. Here we qualify ``not too'' as those
configurations for which we obtain consistent mass profiles using the
two extreme values of $\Upsilon_{in}$, that are allowed for the
extracted value of $\alpha$. Thus, when the method fails, we know that
it does. As long as this aforementioned consistency is noted, choosing
$\Upsilon_{in}$ from anywhere within the range corresponding to the
given $\alpha$ will lead to consistent total mass distributions within
3$X_e$.  Additionally, this range is neither too constricted nor too
relaxed, as was discussed in Paper~I.

\subsection{Comparatively better applicability}
\noindent
The presented device is based upon conclusions that are drawn from a
sample of model Sersic galaxies. This would naturally imply that the
success of this formalism is crucially dependant on the generality of
the models. In particular, we have discussed the ranges of $n$ and
$X_e$ for which our advanced results are true. We have also discussed
the effect of changing properties of the dark halo that we use in the
models and find the advanced scheme robust to such model parameter
variations.

In fact, the formalism
presented above is unique in its scope and structure. Estimates of
total mass distributions in distant elliptical systems are difficult
and therefore rare; the formulae presented herein are therefore
advantageous and could be treated as guides to decipher the total
mass distribution of Sersic galaxies in large surveys.

Most importantly, the advanced methodology is successful within a
severely constricted data domain, compared to any other scheme that
aims to obtain mass distributions in elliptical galaxies. All that the
advanced method demands in terms of data is what is typically
available - surface brightness profile and a measure of central
velocity dispersion. The undemanding nature of our method renders it
applicable even at high redshifts. The simplicity of implementation of
the input data is advantageous in that it allows for the scheme to be
used in an automated way, to obtain mass distributions for large
samples of galaxies. Tricks such as this and \cite{trick_08} exploit
the basic configuration within galaxies and offer novel ways for
characterisation of distant systems.

\begin{acknowledgements}
\noindent
DC is funded by a Royal Society Dorothy Hodgkin Fellowship. BJ
acknowledges the support of a University of Nottingham Summer
Studentship. We thank Sebastian Foucaud for useful discussions that
helped enrich the paper.
\end{acknowledgements}

\end{document}